\documentclass[11pt]{article}
\usepackage{color}
\usepackage{tipa}
\usepackage{pgf}
\usepackage{amssymb}
\usepackage{amsfonts}
\usepackage{graphicx}
\usepackage{amsmath}
\usepackage{txfonts}
\usepackage{mathrsfs}
\usepackage{braket}
\usepackage[top=1in, bottom=1in, left=1in, right=1in]{geometry}
\usepackage{bm}
\usepackage{slashed}
\usepackage{indentfirst}
\usepackage{picinpar}
\usepackage{mdwlist}
\usepackage{verbatim}
\usepackage{braket}
\usepackage{authblk}
\usepackage{setspace}
\usepackage{xparse}
\usepackage{lineno}
\author{ Yanbing Liu and  Andrew A. Houck}
\affil{Department of Electrical Engineering, Princeton University, New Jersey 08544, USA}
\title{Quantum electrodynamics near a photonic band-gap}
\date{\today}

\begin{document}
\maketitle  

%\doublespacing
{\bf 
%Experimental study of circuit quantum electrodynamics (QED) have mostly been focused on the interaction between atoms and discrete photonic modes. 

Photonic crystals provide an extremely powerful toolset for manipulation of optical dispersion and density of states, and have thus been employed for applications from photon generation to quantum sensing with NVs and atoms~\cite{Li2015, Clevenson2015}. The unique control afforded by these media make them a beautiful, if unexplored, playground for strong coupling quantum electrodynamics, where a single, highly nonlinear emitter hybridizes with the bandstructure of the crystal. In this work we demonstrate that such hybridization can create localized “cavity” modes that live within the photonic bandgap, whose localization and spectral properties we explore in detail. We then demonstrate that the coloured vacuum of the photonic crystal can be employed for efficient dissipative state preparation. This work opens exciting prospects for engineering long-range spin models~\cite{Douglas2015, Tudela2015} in the circuit QED architecture, as well as new opportunities for dissipative quantum state engineering.
}

The perturbative effect of a structured vacuum is the renowned Purcell effect which states that the lifetime of an atom in such space will be proportional to the local photonic density of states (DOS) near the atomic transition frequency. In practice, the birth of the photonic crystal, which greatly modifies the vaccuum fluctuations, has enabled the control of spontaneous emission of various emitters such as quantum dots~\cite{Lodahl2004, Fujita2005}, magnons ~\cite{Hoeppe2012} and superconducting qubits~\cite{Nick2015}. However, when an atom is strongly coupled to a photonic crystal, non-perturbative effects become important and significantly enrich the physics. For instance, a single photon bound state has been predicted to emerge within the gap~\cite{John1990}, and spontaneous emission of the atom will thus exhibit Rabi oscillation and light trapping behavior. In contrast to electronic band-gap systems, even multiple photons can be simultaneously localized by a single atom, and the coherent photonic transport within the otherwise forbidden band-gap can have a strongly correlated nature~\cite{Rupasov1996, John1997, Longo2010}. In contrast to a system with discrete cavity modes, which is well described by the single mode or multimode Jaynes-Cummings Hamiltonian~\cite{Yoshie2004, Englund2005, Neereja2015}, a continuous density of states enables the formation of a localized state in the band gap.  While other spin-boson problems with continuous DOS have also been studied experimentally~\cite{Astafiev2010, Hoi2012} or theoretically~\cite{Karyn2012, Goldstein2013} with superconducting circuits, this work explores physics near the band edge, where localized states emerge and reservoir engineering becomes possible.

Light-matter interactions are being actively pursued using cold atoms coupled to optical photonic crystals~\cite{Thompson2013, Kimble2014}, where the study of photonic band edge effects requires a combination of challenging nanostructure fabrication and optical laser trapping. Though impressive progress has been made, atoms are only weakly coupled to photonic crystal waveguides~\cite{Kimble2014}, potentially limiting the physics to the the perturbative regime. In this letter, using a microwave photonic crystal and a superconducting transmon qubit, we are able to reach the strong coupling regime of quantum electrodynamics near a photonic band-gap. Physically, this means that a photon emitted by the qubit near the band-gap will be Bragg reflected and reabsorbed by the qubit multiple times before it tunnels out of the photonic crystal. This dynamical process is characterized by the emergence of spectrally resolvable new polariton states, similar to the well-known vacuum Rabi splitting in cavity QED. We will give a more quantitative definition of strong coupling in the following discussion.

Our device consists of 14 unit cells, each of which contains two coplanar waveguide (CPW) sections with different lengths $\ell$ and impedances $Z$ ($\ell_{lo} = 0.45~\mathrm{mm}, \ell_{hi} = 8~\mathrm{mm}$ and  $Z_{lo}=28~\Omega, Z_{hi} = 125~\Omega$). 
These parameters are chosen so that the band edge is within our measurement window ($4-10~\mathrm{GHz}$) and that the bare photonic crystal has a smooth spectrum. The dispersion relation can be calculated using transfer matrices and is given by 
\begin{eqnarray}
2\cos(\omega_c\ell_{lo}/v_p)\cos(\omega_c\ell_{hi}/v_p) - (\alpha + 1/\alpha)\sin(\omega_c\ell_{lo}/v_p)\sin (\omega_c\ell_{hi}/v_p) = 2\cos(k(\ell_{lo}+\ell_{hi}))
\end{eqnarray}
where $v_p$ is the phase velocity in the waveguide,  $k$ is the Bloch wave vector in the Brillouin zone and $\alpha = Z_{hi}/Z_{lo}$ is the impedance asymmetry. Based on the electric field distribution (Bloch wavefunction), we have purposely placed a transmon qubit in the center of one unit cell in the middle of the device. This makes the qubit optimally coupled to the second photonic band. Consequently, only this band is taken into account and a quadratic dispersion relation $E = \hbar\omega_0 + \alpha (k-k_0)^2$ is further assumed. Our device parameters yield $\omega_0 = 7.7~\mathrm{GHz}$. The complete device image is shown in Fig. 1. It is anchored to the base stage ($15~\mathrm{mK}$) of our dilution refrigerator and connected to a typical $50~\Omega$ measurement chain.  

The Hamiltonian of the whole system can be written as 
\begin{eqnarray}
H = \sum_{k}\hbar\omega_{k}a_{k}^{\dagger}a_{k} + \hbar\frac{\omega_q}{2}\sigma^{z} + \hbar\sum_{k}g_{k}(a^{\dagger}_k\sigma^{-}+a_k\sigma^{+})
\end{eqnarray}
Where $\omega_q, \omega_k$ are the frequencies of the bare qubit and the electromagnetic mode with wave vector $k$.
$a^{\dagger}_k (\sigma^{+})$ and $a_k (\sigma^{-})$ are the mode (qubit) raising and lowering operators. 
We have ignored other dissipation channels of the qubit and have also performed a rotating wave approximation. In the single particle spectrum, there exists a polariton state within the band gap with the eigenenergy $\omega_b$ given by the root of the equation 
\begin{eqnarray}\label{EigenEnergy}
\hbar(\omega_q - \omega_b)\sqrt{\hbar(\omega_0 - \omega_b)} = \pi g^2/\alpha 
\end{eqnarray}
We have already assumed that $g_k \approx g$ for all wave vector $k$, which is valid in our device design. The solution in the band-gap always exists no matter how far the bare qubit frequency $\omega_q$ is detuned from the band edge $\omega_0$. In real space, the photonic part of this polariton state is exponentially localized around the qubit (hence the bound state) with the localization length $L$ given by the penetration depth $L = \sqrt{\alpha/\hbar(\omega_0 - \omega_b)}$. The qubit component of this state can be computed to be $P_{q} = 2(\omega_b-\omega_0)/(3\omega_b-\omega_q-2\omega_0)$, therefore this state is mostly qubit-like deep within the band-gap while it is mostly photon-like close to the band edge. 

In an infinite photonic crystal, this bound state can result in permanent light trapping~\cite{Longo2010} in photonic transport. However, in our finite system the size of which is comparable to $L$, it is a leaky bound state with a finite spectral linewidth $\gamma$. It can be shown that~\cite{Biondi2014} $\gamma$ is proportional to the overlap of this state's wavefunction with the externally coupled waveguide $\gamma \sim e^{-d_0/2L}$, where $d_0$ is the physical length of the device. When probed with a weak signal, this state assists photonic transport within the band-gap; hence, we observe a Lorentzian transmission peak centered at $\omega_b$. As the bare qubit frequency $\omega_q$ is tuned closer to the band edge, the bound state has a larger localization length and thus carries a larger linewidth. 

We measure the bound state linewidth $\gamma$ and exponentially fit the data to the calculated inverse localization length $1/L$ (Fig. 2(b)). This yields the effective device length $d_{fit}$ to be $146~\mathrm{mm}$, in agreement with the length of the whole device $d_0 = 126~\mathrm{mm}$.
To further validate the above theoretical model, we focus on the cases where the bare qubit frequency is completely within the band. In Fig. 2(a), we observe that the bound state peak below the band edge persists while the input signal at the bare qubit frequency is completely reflected due to destructive interference~\cite{Astafiev2010}. 
%The linewidth of the dip is determined by the coupling strength with the extended Bloch modes. 
Now we can extract $\omega_q,\omega_b$ and fit the data to Equation (\ref{EigenEnergy}). Note that when the bare qubit is resonant with the band edge, the predicted energy shift is $\Delta/2\pi= (\omega_b - \omega_0)/2\pi= (\frac{\pi g^2}{\alpha})^{2/3}\frac{1}{h}$ and $P_{q} = 2/3$. We use $\Delta/2\pi$ as the fitting parameter instead of $g$ so that we can then define the strong coupling regime as $\Delta \gg\kappa$, where $\kappa$ characterizes the steepness of the band edge. In our device, the best fit yields $\Delta/2\pi = 250~\mathrm{MHz}$ while $\kappa/2\pi \approx 26~\mathrm{MHz}$ (see supplement). 

%Here we use transfer matrix techinque to numerically simulate the transmission and extract the corresponding linewidth. Interestingly, when the qubit is inside the band-gap, the linewidth is proportinoal to the coupling strength, essentially Purcell effect. However, when the qubit is in the band, the linewidth of the bound state is inversely proportional to the coupling as it is pushed furhter into the band-gap. 

Unlike harmonic defect states, this bound state can be used to control quantum transport within the band-gap. 
%Hence we observe that the homodyne signal of the transmission near the bound state decreases with increasing input power, accompanied by supersplitting (see supplement). 
We achieve this by taking advantage of the anharmonic multilevel structure of the transmon qubit (anharmonicity $\mathrm{E}_c/2\pi = 385~\mathrm{MHz}$).
We tune the bound state deep into the gap $\omega_0 - \omega_b\gg \Delta$, resonantly pump it with Rabi rate $\Omega_p$ and apply another weak tone to probe the transmission. 
In the limit of $\Omega_p\gg \gamma$, we observe in Fig.~3(a) four extra transmission peaks due to the appearance of Rabi sidebands and an Autler-Townes (AT) splitting of $\omega_{12}$ while the bound state peak is strongly suppressed. 
The AT splitting arises due to the Rabi splitting of the $\ket{1}\leftrightarrow\ket{2}$ transition.
Furthermore, the AT splitting doublet has a much larger transmission amplitude than the Rabi sidebands. Similarly in Fig.~3(b), when the pump tone is resonant with the second transition $\omega_p = \omega_{12}$, we only observe the AT splitting of $\omega_{01}$, also known as electromagnetically induced transparency (EIT) of a single atom. Here the EIT effect is revealed in the suppression of transmission at $\omega_{01}$ within the band-gap. 

These observations are attributed to the coupling of the laser-dressed states with the photonic crystal and their different steady state populations. Essentially, a photon can transmit through the band-gap only if the resonant dressed state transition $\ket{\nu, N} \leftrightarrow \ket{\mu, N+1}$ is strongly coupled to the waveguide and not population inverted, $\rho_{\nu \nu} > \rho_{\mu \mu}$, where $\ket{\nu,N}$ or $\ket{\mu,N+1}$ indicate laser-dressed states. 
This result can be approximately arrived at by assuming linear response and using a transfer matrix technique~\cite{ShenPRL2005}. In a dispersionless waveguide, the transmission coefficient of a driven transmon can be simplified as~\cite{Koshino2013},
\begin{eqnarray}
t_{q} = 1 + \eta(\rho_{\nu\nu} - \rho_{\mu\mu})
\label{eqn::transmission}
\end{eqnarray}
Where $\eta$ is a positive quantity of the order of unit. Combining this with transfer matrices of the perodic waveguides takes into account all quantum inference effects associated with the Bragg reflection and allows us to get the total transmission coefficient $t$.
We see that the probe signal is amplified (attenuated) when the population is inverted (not inverted) in a normal waveguide, while the opposite is true within the photonic band-gap. For instance, in Fig.~3(a) near the bound state frequency $\omega_{01}$, when $\Omega \ll \gamma$, equivalent to the undriven case, $t_q \approx 0$ and $r_q \approx -1$ yielding $t \approx 1$. While for $\Omega \gg \gamma$, the corresponding dressed states are almost equally populated and according to Eq.~\ref{eqn::transmission}, $t_q \approx 1$ and $r_q \approx 0$ resulting in $t \approx 0$. These calculations yield good agreement, and show quantum transport within the band-gap can indeed be coherently controlled with an external drive (see supplement). 

Laser-dressed states can even hybridize with the photonic crystal and form doubly dressed states, just as a single photon bound state is formed when a bare qubit is tuned near the band edge. 
Though being a archetypal quantum optics model, analytical treatment of resonance fluorescence near the band edge is not available~\cite{Breuer}. We present an experimental examination of the driven dynamics as we tune the bound state closer to the band edge. In the pump-probe experiment in Fig.~4(a), we observe that when one sideband gets close to the band edge it splits into two resonances, including a peak within the band-gap and a dip within the band. This level splitting ($\sim 90~\mathrm{MHz}$) is weaker than the direct coupling between the qubit and the photonic crystal $\Delta$ (see supplement). 
This spectral information underlies the non-Markovian light emission dynamics, i.e. the emitted light at the sideband can be reflected back by the photonic medium and re-absorbed by the qubit. 

The deep transmission dip around the upper sideband can be interpreted as dressed state cooling which means that the qubit is dynamically pumped into one specific quantum state. 
Ignoring the band edge effect and higher transmon levels, the reduced dynamics of the qubit can be described by the following master equation (see supplement)
\begin{eqnarray}
\frac{\partial \rho}{\partial t} = \frac{\gamma_o}{8}(\tilde{\sigma}^{z}\rho\tilde{\sigma}^{z} - \tilde{\sigma}^{z}\tilde{\sigma}^{z}\rho) + \frac{\gamma_{-}}{8}(\tilde{\sigma}^{+}\rho\tilde{\sigma}^{-} - \tilde{\sigma}^{-}\tilde{\sigma}^{+}\rho) + \frac{\gamma_{+}}{8}(\tilde{\sigma}^{-}\rho\tilde{\sigma}^{+} - \tilde{\sigma}^{+}\tilde{\sigma}^{-}\rho) + h.c. 
\end{eqnarray}
Here the decay rates at Mollow triplets, $\gamma_{0,\pm}$, are proportional to the local photonic DOS. $\tilde{\sigma}^{z,+,-}$ are Pauli matrices in the dressed state $\ket{\pm} = \frac{1}{\sqrt{2}}(\ket{0} \pm \ket{1})$ basis.
%In free space, $\gamma_0 = \gamma_+ = \gamma_-$ while they can be very different in PhC. 
It is clear that the steady state population of the dressed state $\ket{-}$ is $\rho_{--} = \gamma_{+}/(\gamma_{-} + \gamma_{+}) $. As a result, the qubit will be polarized to the $\ket{-}$ state if the upper sideband falls in the photonic band while the lower sideband falls in the band-gap ($\gamma_+ \gg \gamma_-$). 
%In our device, $\gamma_{+}/\gamma_{-} = {\color{red}???}$. 
In the pump-probe experiment shown in Fig. 4(b), we see that the probe signal at the upper sideband can be suppressed by almost an order of magnitude. According to the linear response theory introduced above, this directly indicates dressed state inversion $\rho_{--} \gg \rho_{++}$. However, the inversion is not perfect due to occupation of higher transmon levels. Compared to a previous reservoir engineering scheme in which a similar effect is demonstrated~\cite{Murch2012}, here only one external drive is required and the effective dressed state cooling is completely caused by the strong asymmetry of the DOS in the photonic crystal. This mechanism can be used to stabilize an arbitrary state on the Bloch sphere by detuning the drive from the qubit. 
Future versions of the device can purposely incorporate a defect cavity mode to assist dispersive readout of the qubit state. 
Furthermore, this can be directly generalized to the many-qubit case where a dynamical quantum phase transition would be observable and highly entangled many body states can be stabilized~\cite{John1996, John1997II}. Also, by engineering the coupling of the qubit with multiple bands, a potential wideband dressed state laser and amplifier can be envisioned. 

To conclude, we have experimentally investigated strong coupling quantum electrodynamics in a photonic crystal. We have observed the single photon bound state and have investigated quantum transport phenomena in the linear regime within the photonic band-gap. In the future, low noise amplifiers can be integrated to study quantum correlation effects in coherent multiphoton transport. The concept can be generalized to a three dimensional architecture using machined waveguides. The superconducting qubit can be replaced by collective excitation of spin ensembles in materials like diamond~\cite{Amsuss2011} or yttrium iron garnet~\cite{Zhang2014}. In this way, the ultrastrong regime may be reached. Finally, this provides a platform for studying spin-models with coupling mediated by overlapping photonics bound states\cite{Douglas2015, Tudela2015}, with built in initialization through reservoir engineering.
%Furthermore, the current design makes integration of DC bias through the center-pin straightforward and can be used to study other solid state systems. 
%Finally, the photonic bound states can be used to mediate interactions between multiple qubits, providing a new platform for quantum simulation with superconducting circuits. 
%Strongly correlated quantum states within the band gap\cite{Rupasov1996, John1997} will emerge and can be studied in similar transport measurement. Thus photonic band structure can be used as a correlated photon filter, extending the concept of microwave filter to the quantum regime. 

\section*{Methods}
The photonic crystal was made using standard optical lithography and dry etching techniques from a $200 \mathrm{nm}$ Nb thin film on a $10~\mathrm{mm} \times 10~\mathrm{mm}$ sapphire substrate. 
The pair of Josephson junctions of the transmon qubit were made using Dolan bridge technique and evaporated with aluminum. 

The whole device is packaged in a printed circuit board, wire bonded and anchored at the base plate (15 mK) of our dilution refrigrator. An external solenoid magnet is used to apply magnetic field across the SQUID loop of the qubit. 

\section*{Acknowledgement}
The authors would like to acknowledge D.~Sadri, G.~Zhang, N. M.~Sundaresan, J.~Simon for valuable discussions.
\section*{Funding}
This work was supported by IARPA under contract W911NF-10-1-0324 and the US National Science Foundation through Materials Research Science and Engineering Centers under contract DMR-1420541.

\section*{Author contributions}
Y.~Liu designed the device, performed the measurements and analyzed the data. A.~A.~Houck supervised the whole experiment.

\begin{figure}
\centering
\includegraphics[scale=0.6]{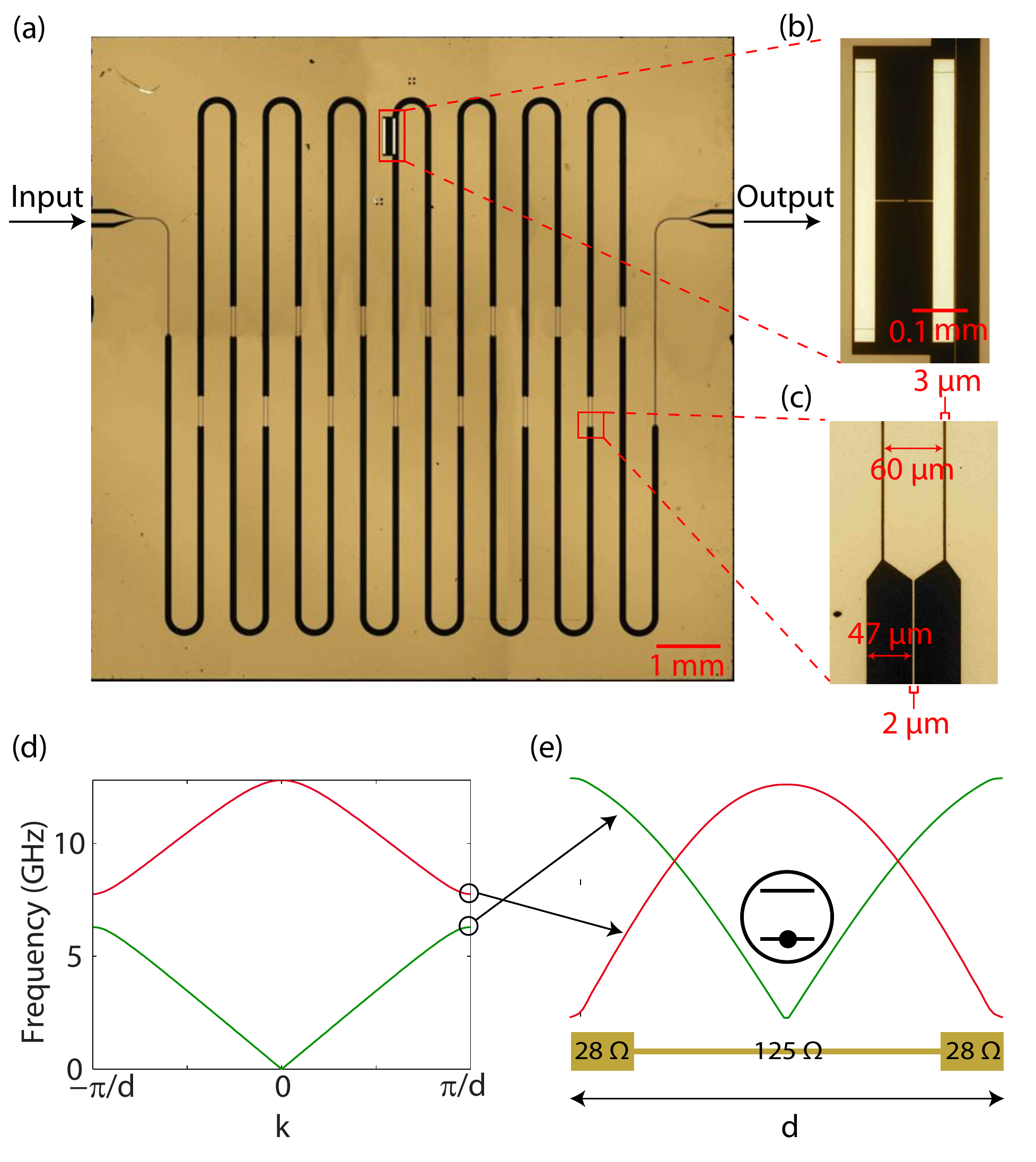}
\caption{Device design. (a) Optical image of the device. The tan region indicates metals Nb and Al while the black region indicates the sapphire substrate. The impedance is periodically modulated by varying the center pin and gap widths of the CPW. (b) A standard transmon qubit made with split pair of Josephson junctions is coupled to the high-impedance section of the unit cell in the middle of the device. (c) The high and low impedance sections of the waveguide. (d) Theoretical band structure of the 1D photonic crystal. 
(e) Schematic of the qubit and the unit cell of the photonic crystal. The green and red curves represent the calculated bloch wavefunction at $k=\pi/d$ of the first and the second photonic band respectively. Placing the qubit in the middle of the unit cell maximizes(minimizes) the coupling with the second(first) band. }
\end{figure}

\begin{figure}
\centering
\includegraphics[scale=0.8]{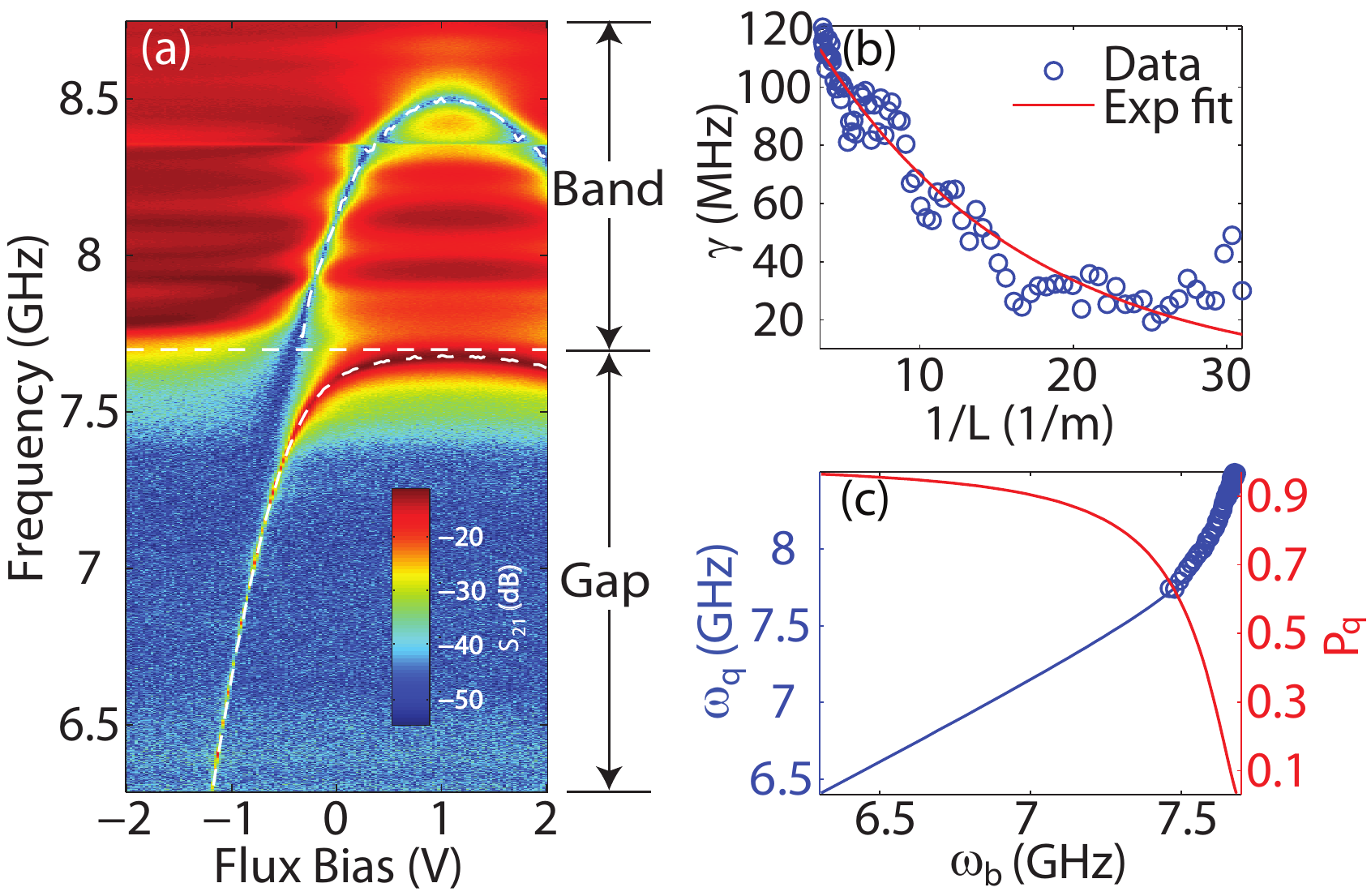}
\caption{Single photon bound state. (a) Low power transmission of the device as the qubit is tuned across the band edge. The colorbar represents the transmission amplitude in log scale. The dashed lines are guides of the eye, indicating the transmission dip within the band (the bare qubit), the peak within the gap (the bound state) and the band edge $\omega_0$ respectively. Note that the defect state at 8.36GHz does not couple to the qubit.
(b) The relation between the bound state linewidth $\gamma$ and the localization length $L$. Based on the theory in the main text, we exponentially fit the data and extract the decay constant to be $146\mathrm{mm}$ which agrees very well with the length of the device ($d_0 = 126\mathrm{mm}$). (c) The relation between the bare qubit frequency and the qubit component with the bound state frequency. The blue curve is extrapolated based on fitting the data when the qubit is within the band.}
\end{figure}

\begin{figure}
\centering
\includegraphics[scale=0.8]{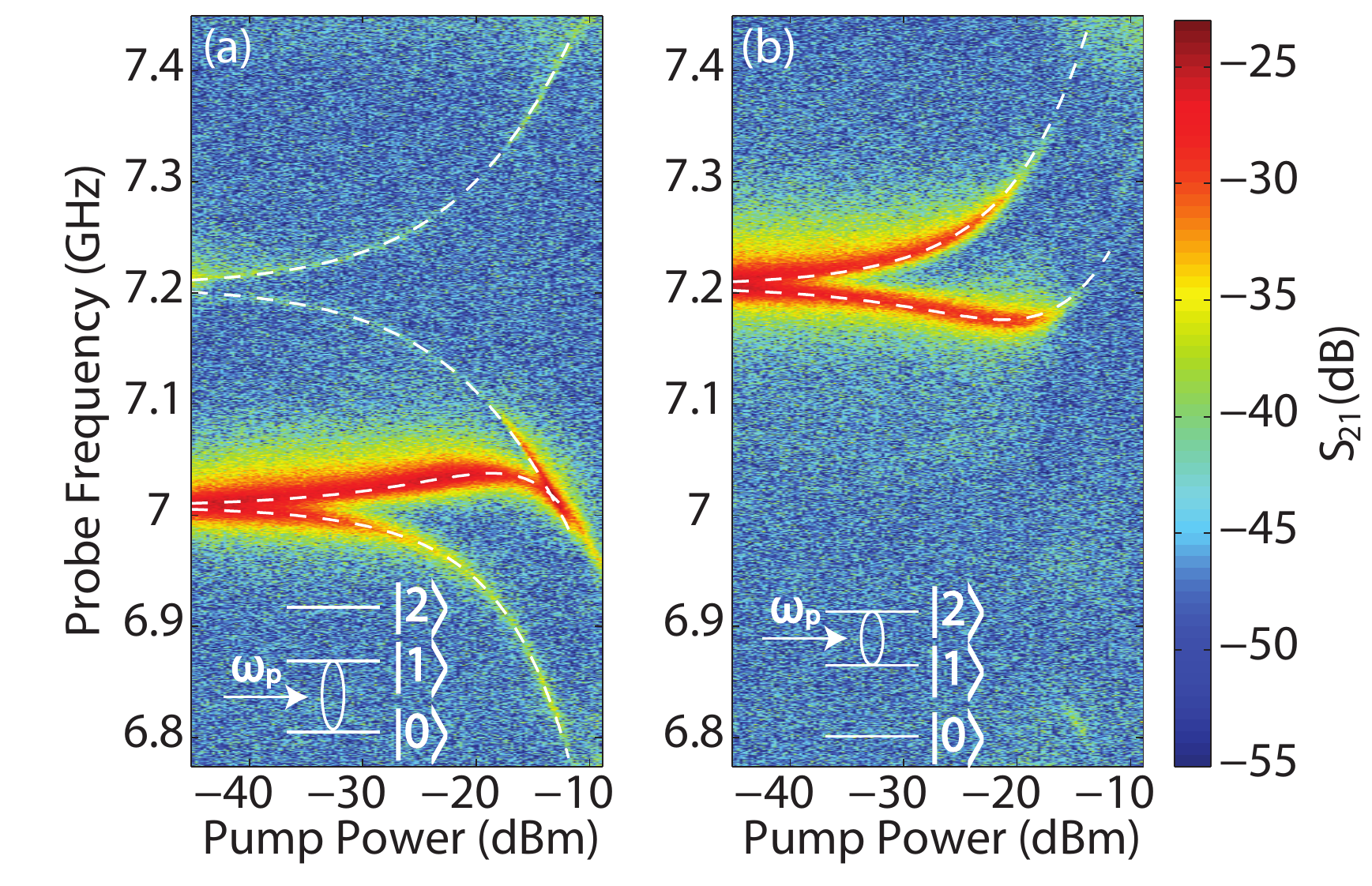}
\caption{Pump-probe experiment within the band-gap. The colorbar represents the transmission amplitude of the weak probe signal in log scale. (a) The bound state is deep within the band-gap ( $\omega_p = \omega_{01} = 7.206\mathrm{GHz}$ and $\omega_{12} = 7.008\mathrm{GHz}$). 
%Note $\omega_{01}-\omega_{12} < E_c$ due to the strongly coupled photonic crystal. 
The transmission near the bound state frequency $\omega_{01}$ is strongly suppressed, accompanied by two Rabi sidebands at approximately $\omega_{01} \pm \Omega$.
The most prominent signature is the Autler-Townes splitting of the second transition near $\omega_{12}$. 
%The transmission peaks and amplitudes manifest different dressed state transitions and their populations. 
%All dressed state transitions are manifested as transmission peaks inside the band-gap. 
Dashed white lines indicate simulation results. Four transmon states are accounted for in the simulation where the only fitting parameter is the effective Rabi rate $\Omega_p$. (b) $\omega_p = \omega_{12}$. In this case, the AT splitting of $\omega_{01}$ is the dominant feature. This phenomenon is also sometimes called single atom EIT.}
\end{figure}

\begin{figure}
\centering
\includegraphics[scale=0.8]{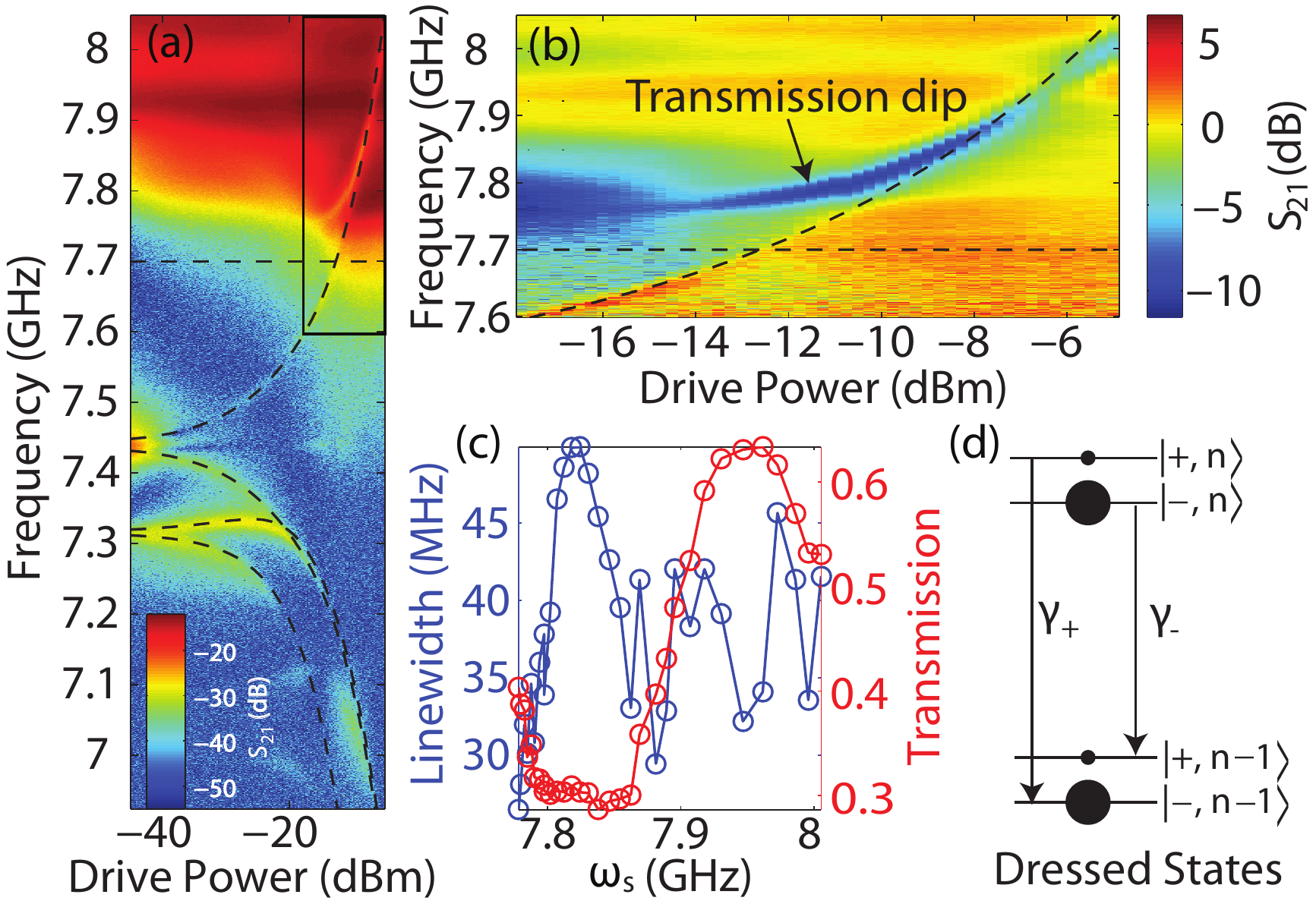}
\caption{Pump-probe experiment near the photonic band edge. (a) One of the sidebands fall within the band. Dashed lines indicate the band edge and dressed state transitions fitted with the same model as in Fig. 3. The top-right boxed region is further shown in (b). The colorbar represents the transmission amplitude of the weak probe signal in log scale. (b) Zoom in around the sideband. We get the new colorbar by subtracting the bare photonic crystal transmission. Note the appearance of additional resonance when the sideband is close to the band edge. The resonance within the band is shown in strong attentuation of the probe signal. 
(c) Extracted linewidth and transmission coefficient based on Fig. 4(b). 
(d) Schematic of the dressed state cooling when $\gamma_+\gg\gamma_-$. }
\end{figure}

\bibliographystyle{unsrt}

\begin{thebibliography}{1}

\bibitem{Li2015} Li,~L. \textit{et al.} Coherent Spin Control of a Nanocavity-Enhanced Qubit in Diamond, \textit{Nat. Commun.} {\bf6}, 6173 (2015).

\bibitem{Clevenson2015} Clevenson,~H. \textit{et al.} Broadband magnetometry and temperature sensing with a light-trapping diamond waveguide,
\textit{Nat. Physics} {\bf11}, 393 (2015).


\bibitem{Douglas2015} Douglas,~J.~S. \textit{et al.} Quantum many-body models with cold atoms coupled to photonic crystals, \textit{Nat. Photon.} {\bf9}, 326 (2015).

\bibitem{Tudela2015} Gonzalez-Tudela,~A., Hung,~C.-L., Chang,~D.~E., Cirac,~J.~I., and Kimble,~H.~J. Subwavelength vacuum lattices and atom-atom interactions in two-dimensional photonic crystals,
\textit{Nat. Photon.} {\bf9}, 320 (2015).

\bibitem{Lodahl2004} Lodahl,~P \textit{et al.} Controlling the dynamics of spontaneous emission from quantum dots by photonic crystal, \textit{Nature} {\bf 430}, 654 (2004).

\bibitem{Fujita2005} Fujita,~M., Takahashi,~S., Tanaka,~Y., Asano,~T. and Noda,~S. Simultaneous inhibition and redistribution of spontaneous light emission in photonic crystals, \textit{Science} {\bf 308}, 1296-1298 (2005). 

\bibitem{Hoeppe2012} Hoeppe,~U. \textit{et al.} Direct Observation of Non-Markovian Radiation Dynamics in 3D Bulk Photonic Crystals, \textit{Phy. Rev. Lett.} {\bf108}, 043603 (2012).

\bibitem{Nick2015} Bronn,~N.~T. \textit{et al.} Broadband filters for abatement of spontaneous emission for superconducting qubits, \textit{Appl. Phys. Lett.} {\bf107}, 172601 (2015). 

\bibitem{John1990} John,~S. and Wang,~J. Quantum electrodynamics near a photonic band gap: Photon bound states and dressed atoms, \textit{Phys. Rev. Lett.} {\bf64}, 2418 (1990).

\bibitem{Rupasov1996} Rupasov,~V.~I. and Singh,~M. Quantum gap solitons and many-polariton-atom bound states in dispersive medium and photonic band gap, \textit{Phys. Rev. Lett.} {\bf77}, 338 (1996).

\bibitem{John1997} John,~S. and Rupasov,~V.~I. Multiphoton localization and propagating quantum gap solitons in a frequency gap medium, \textit{Phys. Rev. Lett.} {\bf79}, 821 (1997).

\bibitem{Longo2010} Longo,~P., Schmitteckert,~P. and Busch,~K. Few-Photon Transport in Low-Dimensional Systems: Interaction-Induced Radiation Trapping, \textit{Phys. Rev. Lett.} {\bf104}, 023602 (2010). 

\bibitem{Burillo2014} Sanchez-Burillo,~E, Zuero,~D, Garcia-Ripoll,~J.~J. and Martin-Moreno,~L. Scattering in the Ultrastrong Regime: Nonlinear Optics with One Photon, \textit{Phys. Rev. Lett.} {\bf113}, 263604 (2014).

\bibitem{John1996} John,~S. and Quang,~T. Quantum Optical Spin-Glass State of Impurity Two-Level Atoms in a Photonic Band Gap, \textit{Phys. Rev. Lett.} {\bf76}, 1320 (1996).

\bibitem{John1997II} John,~S and Quang,~T. Collective Switching and Inversion without Fluctuation of Two-Level Atoms in Confined Photonic systems, \textit{Phys. Rev. Lett.} {\bf78}, 1888 (1997).


\bibitem{Yoshie2004} Yoshie,~T. \textit{et al.} Vacuum Rabi splitting with a single quantum dot in a photonic crystal nanocavity, \textit{Nature} {\bf 432} 200 (2004). 

\bibitem{Englund2005} Englund,~D. \textit{et al.} Controlling the spontaneous emission rate of single quantum dots in a two-dimensional photonic cystal, \textit{Phys. Rev. Lett.} {\bf 95}, 013904 (2005). 

\bibitem{Neereja2015} Sundaresan,~N.~M. \textit{et al.} Beyond strong coupling in a multimode cavity, \textit{Phys. Rev. X} {\bf5}, 021035 (2015).

\bibitem{Astafiev2010} Astafiev,~O. \textit{et al.} Resonance Fluorescence of a Single Artificial Atom, \textit{Science} {\bf327}, 840 (2010).

\bibitem{Hoi2012} Hoi,~I-C. \textit{et al.} Generation of Nonclassical Microwave States Using an Artificial Atom in 1D Open Space, \textit{Phys. Rev. Lett.} {\bf108}, 263601(2012). 

\bibitem{Karyn2012} Le Hur,~K. Kondo resonance of a microwave photon, \textit{Phys. Rev. B} {\bf 85}, 140506(R) (2012).

\bibitem{Goldstein2013} Goldstein,~M., Devoret,~M.~H., Houzet,~M. and Glazman,~L.~I. Inelastic microwave photon scattering off a quantum impurity in a Josephson-junction array, \textit{Phys. Rev. Lett.} {\bf110}, 017002 (2013).

\bibitem{Thompson2013} Thompson,~J.~D. \textit{et al.}, Coupling a single trapped atom to a nanoscale optical cavity, \textit{Science} {\bf340}, 1202 (2013).

\bibitem{Kimble2014} Goban,~A. \textit{et al.} Atom-light interactions in photonic crystal, \textit{Nat. Commun.} {\bf 5}, 3808 (2014).

\bibitem{Biondi2014} Biondi,~M., Schmidt,~S., Blatter,~G. and Tureci,~H.~E. Self-protected polariton states in photonic quantum metamaterials, \textit{Phys. Rev. A} {\bf89}, 025801 (2014). 

\bibitem{ShenPRL2005} Shen,~J.~T. and Fan,~S. Coherent Single-Photon Transport in a One-Dimensional Waveguide Coupled with Superconducting Quantum Bits, \textit{Phys. Rev. Lett.} {\bf95}, 213001 (2005).

\bibitem{Koshino2013} Koshino,~K.  \textit{et al.} Observation of the Three-State Dressed States in Circuit Quantum Electrodynamics, \textit{Phys. Rev. Lett.} {\bf110}, 263601 (2013).
 
\bibitem{Breuer} Breuer,~H.~P. and Petruccione,~F. \textit{The Theory of Open Quantum Systems} (Oxford University Press, Oxford, 2007).

\bibitem{Murch2012} Murch,~K.~W. \textit{et al.} Cavity-assisted quantum bath engineering, \textit{Phys. Rev. Lett.} {\bf109}, 183602 (2012).

\bibitem{Amsuss2011} Amsuss,~R. \textit{et al.} Cavity QED with magnetically coupled collective spin states, \textit{Phys. Rev. Lett.} {\bf107}, 060502 (2011).

\bibitem{Zhang2014} Zhang,~X., Zou,~C., Jiang,~L. and Tang,~H.~X. Strongly coupled magnons and cavity microwave photons, \textit{Phys. Rev. Lett.} {\bf113}, 156401 (2014).
 

\end{thebibliography}

\begin{thebibliography}{1}
\bibitem{ShenPRL2005} J.~T.~Shen and S.~Fan, Coherent Single-Photon Transport in a One-Dimensional Waveguide Coupled with Superconducting Quantum Bits, Phys. Rev. Lett. {\bf95}, 213001 (2005).

\bibitem{John1997}Sajeev John and Tran Quang, Collective Switching and Inversion without Fluctuation of Two-Level Atoms in Confined Photonic systems, Phys. Rev. Lett. {\bf78}, 1888 (1997).

\bibitem{Calajo2015} G.~Calajo, F.~Ciccarello, D.~Chang and P.~Rabl, Atom-field dressed states in slow-light waveguide QED, arXiv:1512.04946.

\bibitem{Shi2015} Tao Shi, Ying-Hai Wu, A.~Gonzalez-Tudela, J.~I.~Cirac, Bound states in boson impurity models, arXiv:1512.07238.

\end{thebibliography}

\clearpage
%\part{}

\title{\huge \centering Supplementary material \\}
%\author{\normalfont Yanbing Liu and  Andrew A. Houck}

\maketitle

\section{Single photon bound state calculation}
Here we show the derivation of the eigenenergy of the single photon bound state within the band gap. The Hamiltonian of our system is 
\begin{eqnarray}
H = \sum_{k}\hbar\omega_{k}a_{k}^{\dagger}a_{k} + \hbar\frac{\omega_q}{2}\sigma^{z} + \hbar\sum_{k}g_{k}(a^{\dagger}_k\sigma^{-}+a_k\sigma^{+})
\end{eqnarray}
The number of excitation $\hat{N} = \sigma_z/2 + \int_{B.Z.}a^{\dagger}_ka_k$ is a conserved quantity in our model. So in the single-excitation manifold, the eigenstate is in the form of $\ket{\phi_b} = \cos\theta\ket{0}\ket{e} + \sin\theta\int_{B.Z.}dkc_ka^{\dagger}_k\ket{0}\ket{g}$ and the eigenstate equation is $H\ket{\phi_b} = \hbar\omega_b\ket{\phi_b}$. This yields
%\begin{eqnarray}
%\omega_a\cos\theta + \sin\theta\int dkg_kc_k = E_b\cos\theta \\
%\omega_k\sin\theta c_k + \cos\theta g_k = \sin\theta c_k E_b
%\end{eqnarray}
\begin{eqnarray}
\hbar(\omega_b - \omega_a) = \int_{B.Z.} dk\frac{\hbar^2g_k^2}{E_b - \hbar\omega_k} \\
\tan^2(\theta) = \int_{B.Z.} dk \frac{\hbar^2g^2_k}{(\hbar\omega_b - \hbar\omega_k)^2}\\
c_k = \frac{g_k}{\tan\theta(\omega_b - \omega_k)}
\end{eqnarray}
To get an analytical result, we further make two assumptions: the coupling is independant of wavevector $g_k \approx g$ and the dispersion is quadratic $\hbar\omega_k = \hbar\omega_0 + \alpha(k-k_0)^2$. Then extending the integral limits to infinity and performing the integrals yield the following results:
\begin{eqnarray}
\hbar(\omega_q - \omega_b) = \frac{\pi g^2}{\alpha\sqrt{\hbar(\omega_0-\omega_b)}} \\
\tan^2(\theta) = \frac{\omega_q - \omega_b}{2(\omega_0 - \omega_b)}
\end{eqnarray}
Thus, the photonic part of the wavefunction $\ket{\phi_b}$ is $\int_{B.Z.}dk\frac{g}{\tan\theta(\omega_b-\omega_k)}a^{\dagger}_k\ket{0}\ket{g}$. By performing a fourier transform $a_k \rightarrow a_x$ and approximate the Bloch wavefunction $\psi_k(x)e^{ikx}$ with $\psi_{k0}(x)e^{ikx}$, we get the photonic part of the wavefunction is in the following form: 
\begin{eqnarray}
\int dx e^{-x/\lambda}a^{\dagger}_x\ket{0}\ket{g} 
\end{eqnarray}
where $\lambda$ is the penetration depth defined as $\lambda = \sqrt{\alpha/\hbar(\omega_0 - \omega_b)}$. The means that photonic part of this polarition state is exponentially localized around the qubit, hence this is often called single photon bound state. 

\section{Band structure and tranfer matrix technique}
We first analyze the mode structure (Bloch wavefunction) of an infinite microwave photonic crystal (PhC). 
It can be obtained by solving the following wave equations for the periodic structure. This calculation leads to the choice of the qubit's position, the midde of the unit cell, for maximal coupling with the second band. 
\begin{eqnarray}
\frac{d}{dx}V(x,t) = -l(x)\frac{d}{dt}I(x,t) \\ \nonumber
\frac{d}{dx}I(x,t) = -c(x)\frac{d}{dt}V(x,t)
\end{eqnarray}
Here $l(x), c(x)$ are the inductance and capacitance per unit length, leading to the following wave equation
\begin{eqnarray}
\frac{\partial}{\partial x}[\frac{v_p}{Z_c(x)}\frac{\partial V(x,t)}{\partial x}] = \frac{1}{v_pZ_c(x)}\frac{\partial^2}{\partial t^2}V(x,t)
\end{eqnarray}
Here $v_p = \frac{1}{\sqrt{lc}}$ is the phase velocity and $Z_c(x) = \sqrt{\frac{l(x)}{c(x)}}$ is the characteristic impedance. $Z_c(x)$ is periodically modulated by changing the center pin and gap widths of the coplanar waveguide. This 1D wave equation can be solved using standard fourier transform technique assuming $V_k(x) = \Sigma_nC_n(k)e^{i2n\pi x/d+ikx}$ and $1/Z_c(x) = \Sigma_m\eta_me^{i2m\pi x/d}$. Then the algebraic equation can be solved numerically. The results are shown in Fig. S1.

\begin{figure}
\centering
\includegraphics[scale=0.8]{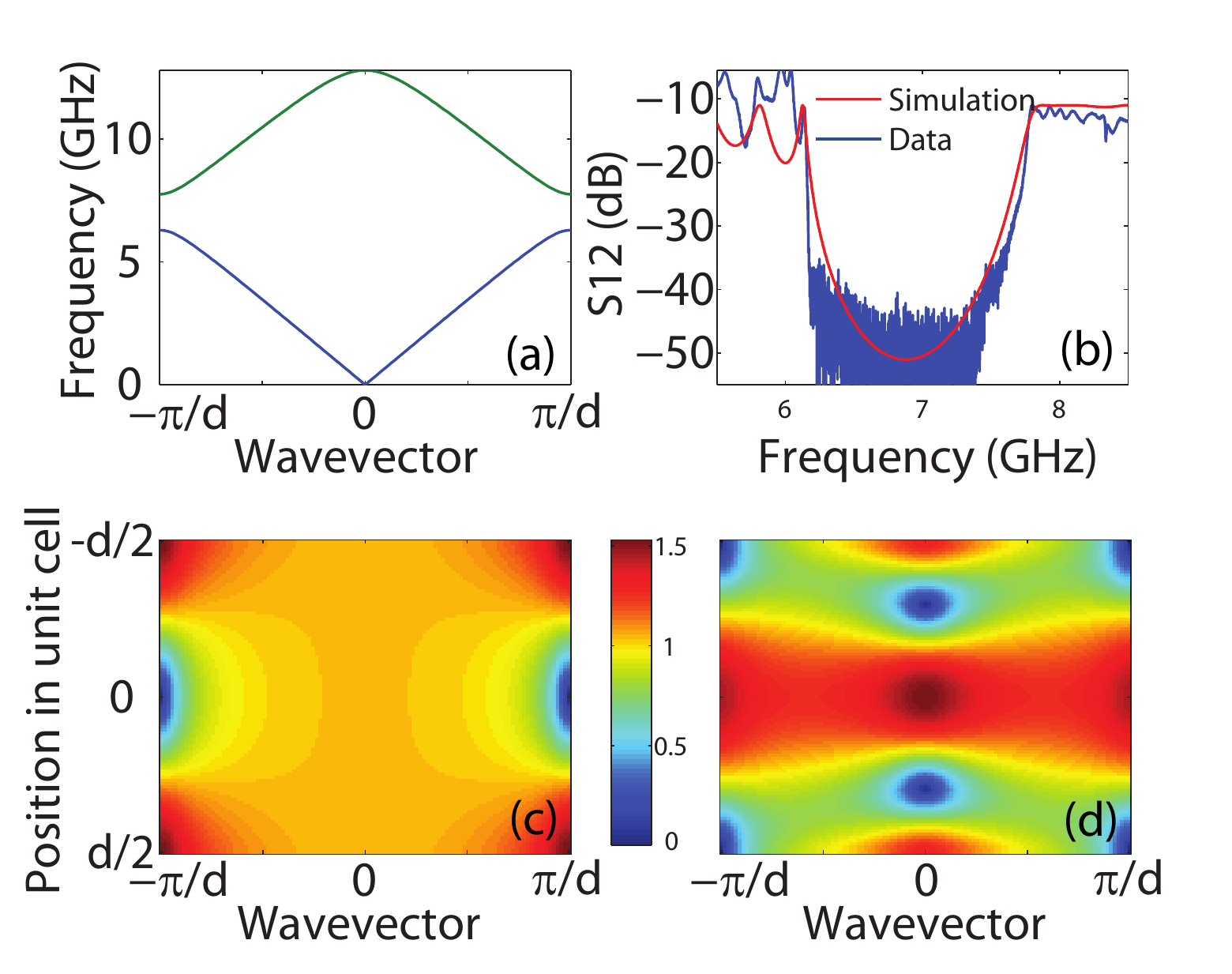}
\caption{Band structure of the bare photonic crystal. (a) Dispersion of the PhC. (b) Measured (at 10mK) and simulated (with ABCD matrix technique) transmission amplitude in the log scale.
The small discrepancy comes from fabrication imperfection, additional loss and slight impedance mismatch in the whole measurement chain. Note there are no pronoused resonances in the second band. Mode structure (Bloch wavefunction) of the first (c) and second band (d). The most strongly-coupled modes, which lie near the band-gap, have wavevector $k=\pi/d$. In the middle of the unit cell, clearly the mode in the first band reaches minimum while the mode in the second band reaches maximum.
}
\end{figure}

\begin{figure}
\centering
\includegraphics[scale=0.5]{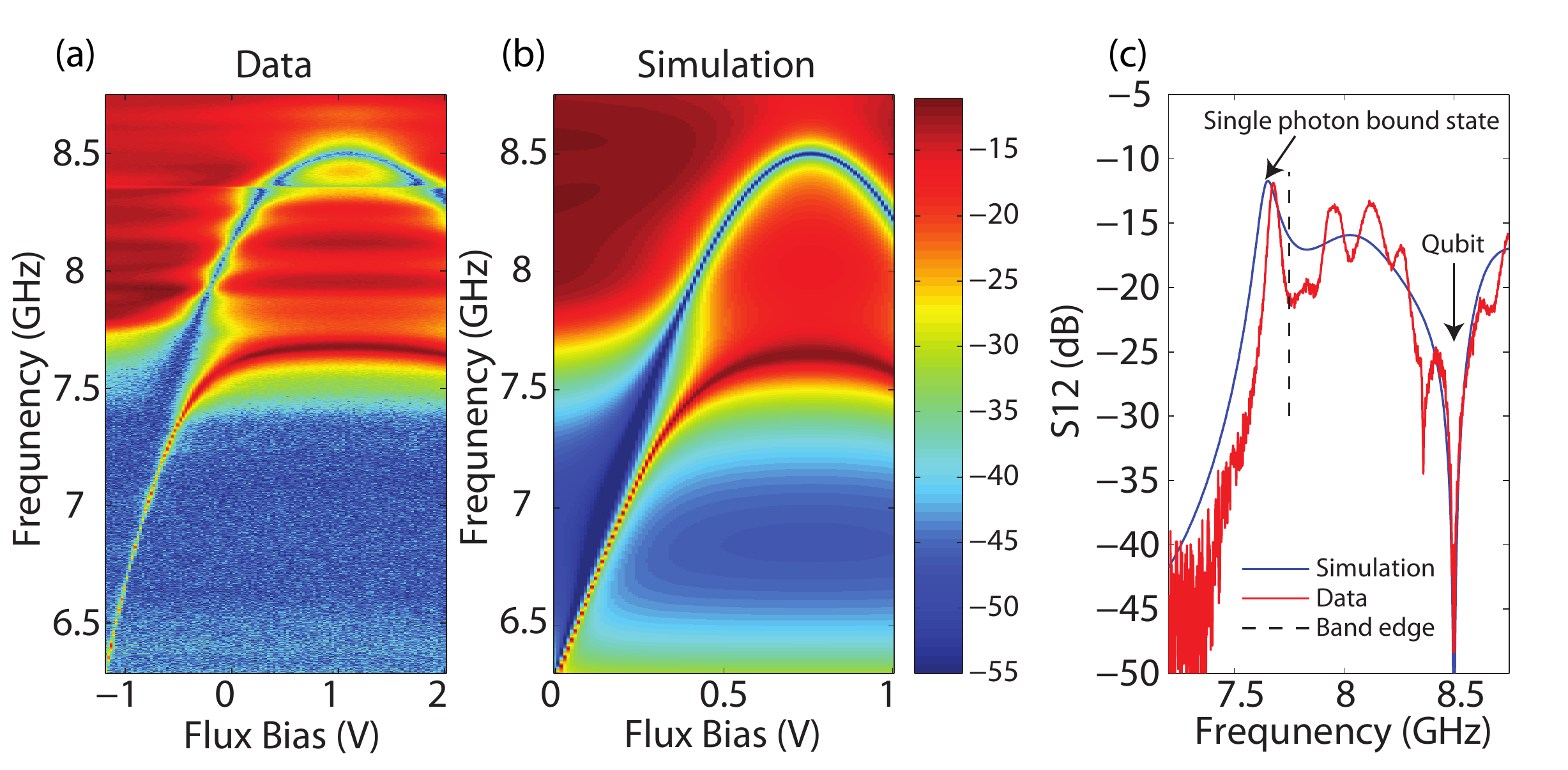}
\caption{ABCD matrix simulation. (a) Measured low-power transmisson versus flux bias in log scale. (b) Simulated transmission with ABCD matrix technique in log scale. Here we have assumed $\gamma/2\pi = 0.5\mathrm{GHz}$. This calculation gives qualitative agreement. But we only use the Hamiltonian approach to fit the experimental data.
(c) S21 at $\omega_a/2\pi = 8.5\mathrm{GHz}$. } 
\end{figure}

To simulate the transmission of our device with a finite number of periods, the transfer matrix (also called ABCD matrix in microwave engineering literature) technique can be employed. The ABCD matrix of a section of coplanar waveguide with characteristic impedance $Z$ can be written as 
\begin{eqnarray}
M_{Z} = 
\begin{bmatrix}
cos(\omega \ell/v_p) &  jZsin(\omega \ell/v_p)\\
jsin(\omega \ell/v_p)/Z & cos(\omega \ell/v_p)
\end{bmatrix}
\end{eqnarray}
Then the ABCD matrix for one unit cell is $M_{unit} = M_{Z_{hi}}M_{Z_{lo}}$. 

The ABCD matrix for an atom in the single photon regime\cite{ShenPRL2005} is given by:
\begin{eqnarray}
M_a = 
\begin{bmatrix}
1 &  0\\
-j\frac{\gamma(\omega_a)}{\omega - \omega_a}\frac{1}{Z_0} & 1
\end{bmatrix}
\end{eqnarray}
Here $\gamma(\omega_a)$ is the decay constant in a normal waveguide with characteristic impedance $Z_c$. This constant is set by the coupling strength between the qubit and the waveguide. To maximize this coupling, one capacitor island of the transmon qubit is fabricated very close ($2\mu \mathrm{m}$) to the center pin of the waveguide. In the simulation shown in Fig. S2, we have assumed that $\gamma(\omega_a)$ is constant for all $
\omega_a$.

The ABCD matrix for the whole device can then be written as $M = (M_{unit})^N\times M_a\times (M_{unit})^N$
and the transmission coefficient can then derived. The explicit expression is too cumbersome to be presented here, so instead we give the numerical results.

\section{Nonlinear response and comments on multiphoton bound states}
Neither of the previous theoretical approaches can be easily generalized to the nonlienar regime. 
A few very recent theoretical studies have used variational ansatz and numerical methods to prove the existence of multi-photon bound states\cite{Calajo2015, Shi2015}. We did not observe a direct signature of these multiphoton bound states in our experiments. As we increase input power, the bound state assisted transmission coefficient decreases accompanying with supersplitting. The supersplitting doublet appears due to saturation of the two-level system and does not correspond to multiphoton transitions. We think that the absence of multiphoton resonances can be attributed to insufficiently strong coupling and relatively large bandwidth of the single photon bound state. The Mollow triplet structure and Autler-Townes splitting clearly shows the quantum nature of this state within the band-gap. Incorporation of low noise amplifiers in the future may reveal multitphoton effects in correlation measurements. 

\section{Multilevel analysis of dressed-state transitions}
The Hamiltonian of the qubit subject to coherent drive has the following form,
\begin{eqnarray}
H = \sum_{n=0,1...}[(\omega_n - n\omega_d)\ket{n}\bra{n} + \frac{\Omega_n}{2}(\ket{n}\bra{n+1} + h.c.)]
\end{eqnarray}
The above hamiltonian is written in the frame of the drive and multiple excited states are taken into account. Where $\Omega_n = \sqrt{\gamma_n}E$ is the effective Rabi rate for the nth transition and Fermi's golden rule dictates that $\sqrt{\gamma_n}$ is proportional to the transition matrix of transmon qubit. Note that the energy levels here have taken into account the effect of the photonic crystal, so $\omega_{01}$ is really just the bound state frequency $\omega_b$. The other transition frequencies $\omega_{12}, \omega_{23}$ are found empirically in the experiments, rather than using the transmon anharmonicity $E_c$.
Hence $\Omega_n = \sqrt{n+1}\Omega_0$. We note that the full quantum analysis has to use input-output theory and consider the coherent coupling between the qubit and photonic crystal modes. However, if the qubit is deep within the gap, the effect of the photonic modes can be described perturbatively with the decay rate $\gamma_n$. Diagonizing the matrix involving a few energy levels yields the complete dressed states. We consider two experimentally relevant cases where $\omega_d = \omega_1 - \omega_0$ and $\omega_d = \omega_2 - \omega_1$. These dressed states are fitted to the experimental data. The only fitting parameter is $\Omega_0$.
The dressed state transition shown in Figure 3 in the main text thus correspond to $\ket{a} \leftrightarrow \ket{b}$ for $\omega_d = \omega_{01}$. and for $\omega_d = \omega_{12}$.

\section{Effective master equation in photonic crystal}
To understand resonance fluorescenc in the band-gap medium, we first start from the complete Hamiltonian including both the coherently-driven qubit and the modes of the photonic crystal.
\begin{eqnarray}
H = \frac{\Delta_a}{2}\sigma_z + \frac{\Omega_0}{2}(\sigma^{+}+\sigma^{-}) + \sum_{k}\Delta_{k}a^{\dagger}_ka_k + 
\sum_kg_{k}(a^{\dagger}_k\sigma^-+a_k\sigma^+) 
\end{eqnarray}
Where $\Delta_a = \omega_a-\omega_d, \Delta_k = \omega_k - \omega_d$. 
We first go to the dressed state of the qubit, defining $\ket{\tilde{0}} = \cos(\theta)\ket{0} - \sin(\theta)\ket{1}, \ket{\tilde{1}} = \sin(\theta)\ket{0} + \cos(\theta)\ket{1}$, where
$\cos^2(\theta) = \frac{1}{2} + \frac{\Delta_a}{2\sqrt{\Omega_0^2+\Delta_a^2}}$.

\begin{eqnarray}
H =  \frac{\Omega}{2}\tilde{\sigma}_z + \sum_{k}\Delta_{k}a^{\dagger}_ka_k + 
\sum_kg_{k}(\cos^2(\theta)a^{\dagger}_k\tilde{\sigma}^- - \sin^2(\theta)a^{\dagger}_k\tilde{\sigma}^+ + \sin(\theta)\cos(\theta)a^{\dagger}_k\tilde{\sigma}^z) 
\end{eqnarray}
The Rabi rate is given by $\Omega = \sqrt{\Omega_0^2 + \Delta_a^2}$
We work in the rotating frame, applying a unitary transformation $U=\exp[i(\Omega\tilde{\sigma}_z/2+i\sum_k\Delta_ka^{\dagger}_ka_k)t]$:
\begin{eqnarray}
H(t) = \sum_kg_{k}(\cos^2(\theta)a^{\dagger}_k\tilde{\sigma}^-e^{i(\Delta_k-\Omega)t} - \sin^2(\theta)a^{\dagger}_k\tilde{\sigma}^+e^{i(\Delta_k+\Omega)t} + \sin(\theta)\cos(\theta)e^{i\Delta_kt}a^{\dagger}_k\tilde{\sigma}^z) 
\end{eqnarray}
From the above Hamiltonian, we can follow the standard procedure involving Born-Markov approximation and rotating wave approximation, deriving the reduced master equation for the qubit. But before we do that, let us analyze it to get some physical intuition. If we assume that the upper sideband is close to the band edge $\omega_0 \approx \omega_d + \Omega$, then the effective interaction strength will be $g_k\cos^2(\theta)$, and the coupling of the sideband will be effectively smaller than the direct coupling between the qubit and the photonic modes. The reduced density matrix of the qubit is govened by the following equation, 

\begin{eqnarray}
\frac{d\rho}{dt} = - \int_0^{t}d\tau Tr_R{[H(t), [H(\tau), \rho(\tau) \otimes \rho_R(\tau)]]}
\end{eqnarray}
We assume $\rho(\tau)\rightarrow \rho(t)$ and $\rho_R(\tau)\rightarrow \rho_R(0)$(Born-Markov approximation), then we can get
\begin{eqnarray}
2\frac{\partial \rho}{\partial t} = \gamma_o\sin^2(\theta)\cos^2(\theta)(\tilde{\sigma}^{z}\rho\tilde{\sigma}^{z} - \tilde{\sigma}^{z}\tilde{\sigma}^{z}\rho) + \gamma_{-}\sin^4(\theta)(\tilde{\sigma}^{+}\rho\tilde{\sigma}^{-} - \tilde{\sigma}^{-}\tilde{\sigma}^{+}\rho) \\ \nonumber 
+ \gamma_{+}\cos^4(\theta)(\tilde{\sigma}^{-}\rho\tilde{\sigma}^{+} - \tilde{\sigma}^{+}\tilde{\sigma}^{-}\rho) + h.c. 
\end{eqnarray}
Where $\gamma_0 = 2\pi\sum_kg_k^2\delta(\omega_k-\omega_L)$, $\gamma_{-} =2\pi\sum_kg_k^2\delta(\omega_k-\omega_L+\Omega)$ and $\gamma_{+} = 2\pi\sum_kg_k^2\delta(\omega_k-\omega_L-\Omega)$. 
It's easy then to get the steady state of the master equation and get
\begin{eqnarray}
\bra{\tilde{1}}\rho\ket{\tilde{1}} = \frac{\gamma_{-}\sin^4(\theta)}{\gamma_{-}\sin^4(\theta) + \gamma_{+}\cos^4(\theta)} \\
\bra{\tilde{0}}\rho\ket{\tilde{0}} = \frac{\gamma_{+}\cos^4(\theta)}{\gamma_{-}\sin^4(\theta) + \gamma_{+}\cos^4(\theta)} 
\end{eqnarray}
If the dephasing is also included, then the above equations shall be modified to be 
\begin{eqnarray}
\bra{\tilde{1}}\rho\ket{\tilde{1}} = \frac{\gamma_{-}\sin^4(\theta)+\gamma_p\sin^2(2\theta)}{\gamma_{-}\sin^4(\theta) + \gamma_{+}\cos^4(\theta)+2\gamma_p\sin^2(2\theta)} \\
\bra{\tilde{0}}\rho\ket{\tilde{0}} = \frac{\gamma_{+}\cos^4(\theta)+\gamma_p\sin^2(2\theta)}{\gamma_{-}\sin^4(\theta) + \gamma_{+}\cos^4(\theta)+2\gamma_p\sin^2(2\theta)} 
\end{eqnarray}
In the case of resonant driving $\theta=\pi/4$, the target dressed state is simply $\frac{1}{\sqrt{2}}(\ket{0}-
\ket{1})$. The results can be generalized to many atoms\cite{John1997}.

\end{document}